\documentclass[%
 aip,
 jmp,%
 amsmath,amssymb,
 reprint,%
]{revtex4-2}

\usepackage{graphicx}
\usepackage{dcolumn}
\usepackage{bm}
\usepackage{lmodern}
\usepackage{caption}
\usepackage{subcaption}
\usepackage{float}

\begin{document}

\preprint{AIP/123-QED}

\title{Saddle-node bifurcation during the relaminarization of turbulent puffs in pipe}
\author{Basheer A. Khan}%
 \altaffiliation[Graduated from the Indian Institute of Technology (I.I.T.) Kanpur. Currently a postdoctoral fellow at the ]{Department of Mechanical Engineering, Ben-Gurion University, Israel.}
\author{Shai Arogeti}%
\author{Oriel Shoshani}%
\author{Alexander Yakhot}
\altaffiliation[Corresponding author, yakhot@bgu.ac.il ]{}%
\affiliation{
Department of Mechanical Engineering, Ben-Gurion University, Israel
}%

\begin{abstract}
Turbulent puffs in a pipe persist for a long time before abruptly transitioning to laminar flow through viscous exponential decay. Direct numerical simulation results reveal a saddle-node bifurcation sequence governing the final relaminarization stage. Specifically, a universal square-root scaling law in the annihilation of critical points that is indicative of a saddle-node bifurcation. The Riccati equation has been proposed as a straightforward approach to modeling these bifurcations.
\end{abstract}
\maketitle
\begin{figure}
\centerline{
  \hbox{
    \resizebox{70mm}{!}
    {\includegraphics[width=1.0 \textwidth]{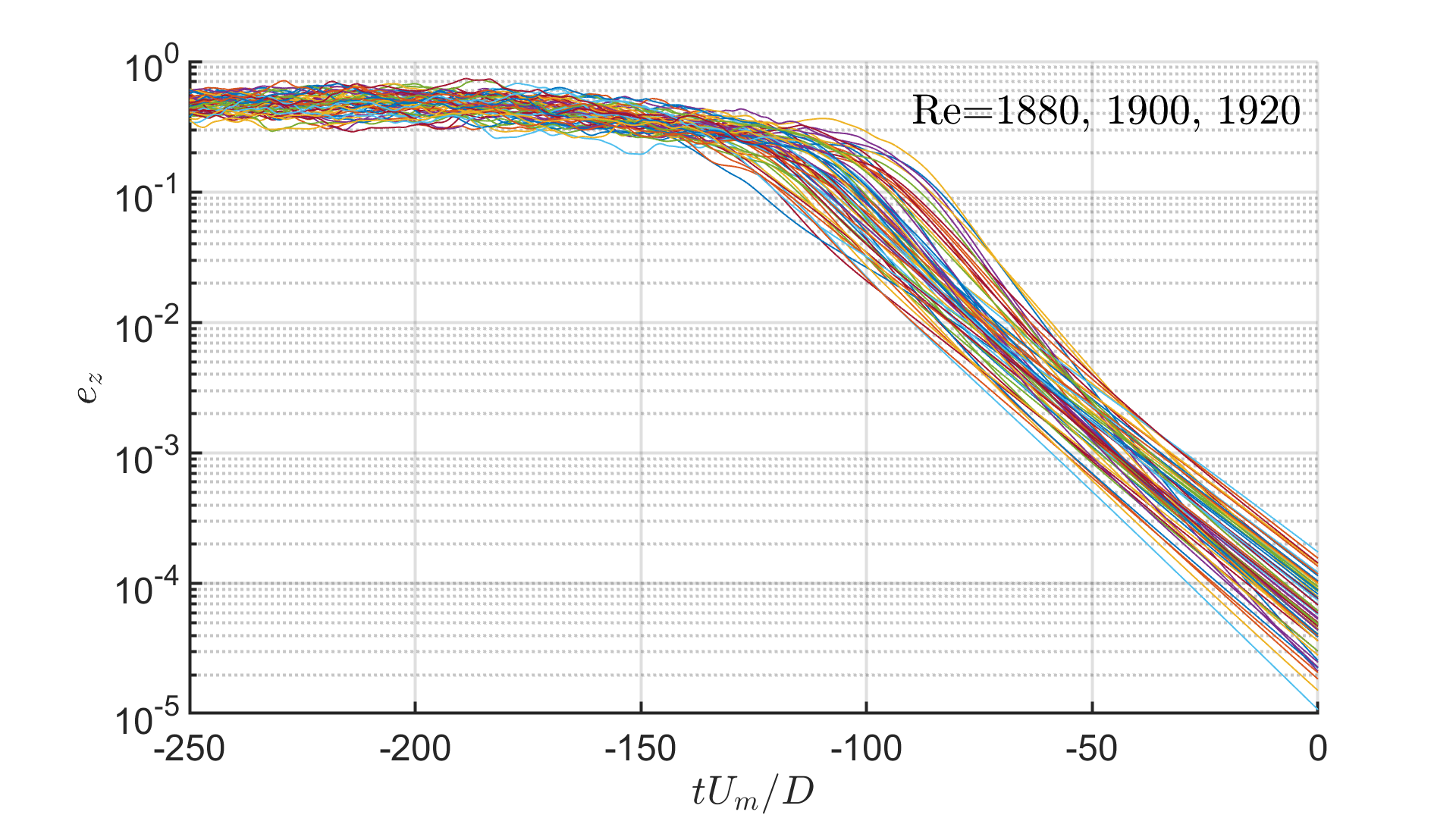}} 
    \hspace{1mm}
    \resizebox{70mm}{!}
    {\includegraphics[width=1.0 \textwidth]{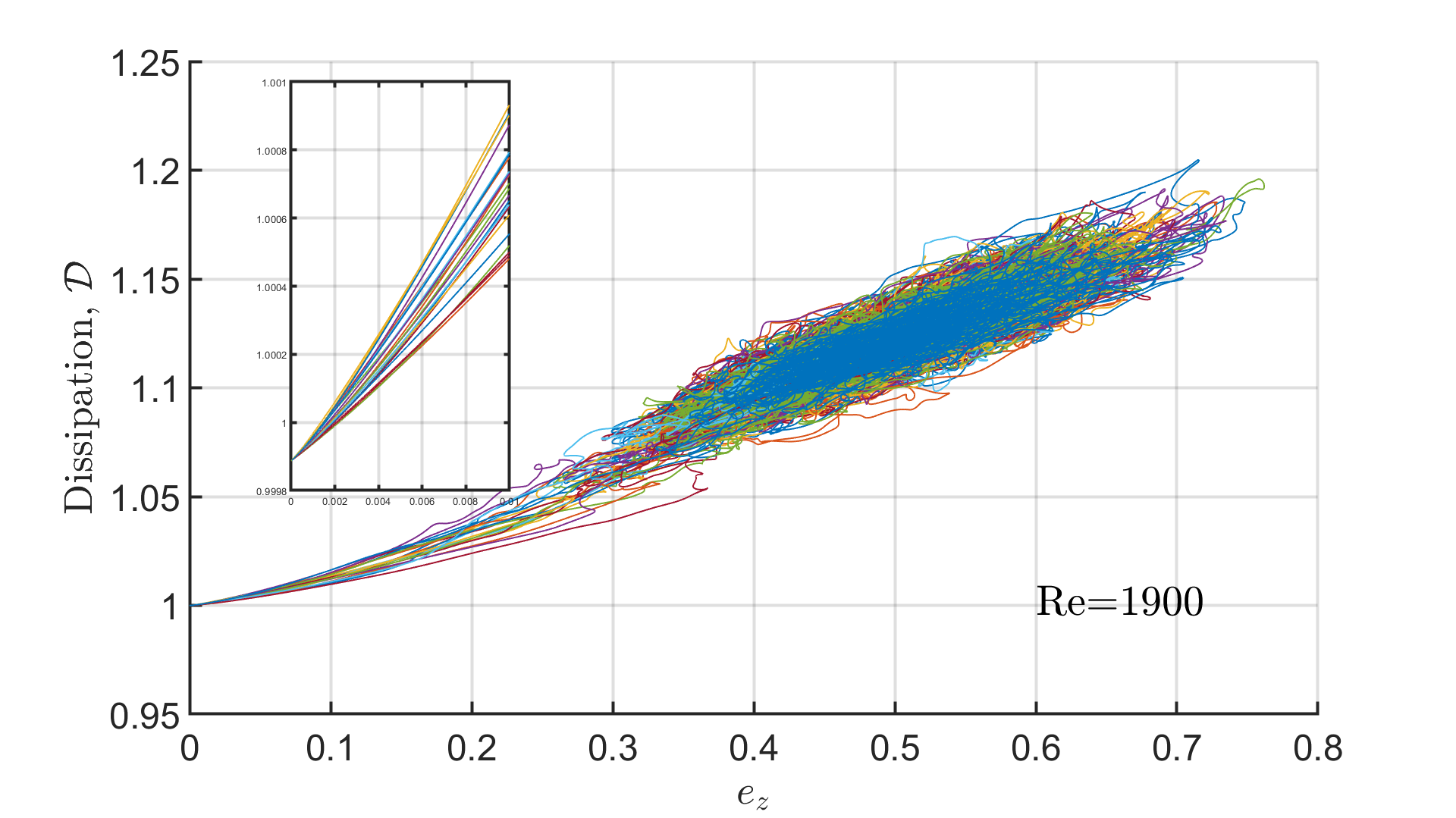}}    
    \hspace{1mm}
  }
}
    \caption{(left)
The last 250 advective time units $(D/U_m)$ before complete relaminarization at t=0; 22 initial conditions (IC) for each $Re$.
(right)  Typical for all Reynolds numbers, phase-space projection of chaotic trajectories (including the relaminarization stage) of the energy of streamwise velocity fluctuations, $e_z$, and the total energy dissipation, $\cal {D}$, computed for different initial conditions (IC).}
        \label{fig: ez-t-ez-vs-D-250last-Re1880-1900-1920}
\end{figure}

\date{\today}
%
%
%
Turbulence in a pipe appears as ``puffs",  which are localized, self-sustaining, and confined patches\cite{Wygi1973} that
occur at Reynolds numbers between 1700 and 2000 (see references \onlinecite{Barkley2016, Avila2023}, and references
therein). Thirty years after turbulent puffs were first found, it was discovered that they don't last forever. Extensive measurements\cite{Hof2006,Peixinho2006} and numerical simulations\cite{Willis2007, Avila2011} showed that puffs survive for a long time before decaying as the Reynolds number increases. Puff splitting was an additional concern requiring thorough investigation as the Reynolds number increases.
Studies of the critical Reynolds number at which the splitting occurs were resolved by the findings of reference \onlinecite{Avila2011}. Measurements and direct numerical simulation (DNS) results were approximated by a single superexponential fit,  $\tau=\exp[\exp(aRe+b)]$,
for the mean equilibrium lifetime of a puff before sudden decaying or splitting. This indicates
approximately 2040 as a critical $Re$-limit for the most probable occurrence of a localized puff in a pipe.
Below the critical point $Re \approx 2040$, the decay is characterized by a mean lifetime that depends on $Re$ and is greater than the mean lifetime before the decaying event\cite{Avila2011}. In particular, the
equilibrium lifetime of turbulent puffs varies from a few hundred to $\sim 2\times 10^7$ advective time units ($D/U_m$) in the
Reynolds number range $Re$=1720--2040. \\
\indent In this Letter we report the results of direct numerical simulations (DNS) of laminar-turbulent transitional flow in a pipe. We used the Openpipeflow\cite[]{Willis2017} Navier-Stokes solver, the Reynolds number range $Re$=1720—1920, $Re=U_mD/\nu$, the bulk velocity $U_m$  was fixed (for details, Refs. \onlinecite{ KhanArogetiYakhot2024} and \onlinecite{KhanArogetiYakhotPoF}). The data were collected inside an $8D$-width window centered around $z=0$, where the instantaneous transverse motion energy $e_{\bot}=\int\int(u_r^2+u_{\theta}^2)r\text{d}r\text{d}\theta$ was found to be the highest; hereafter, $z$ denotes the location in the reference frame linked to the moving puff. At $z = -8D$, the laminar parabolic velocity profile remains nearly undisturbed; however, at $z = -4D$, the onset of turbulence was found to be evident from a slight decrease in centerline velocity and the appearance of an inflection point in the velocity profile, consistent with the observations of reference \onlinecite{Hof2010}. Conventionally, we refer to $z=-4D$ as the onset location of the transition.
Figure \ref{fig: ez-t-ez-vs-D-250last-Re1880-1900-1920} (left) shows the temporal evolution of the energy of longitudinal velocity fluctuations for 22 initial conditions for each of the three Reynolds numbers. Here, 66 curves are aligned by decay time, a countdown from t=0, indicating complete relaminarization for each curve.
As a heuristic threshold in the Openpipeflow code that we used, a flow is considered
to be fully relaminarized when the centerline volume-average velocity is greater than 0.999. The curves in figure
\ref{fig: ez-t-ez-vs-D-250last-Re1880-1900-1920} (left) visually confirm the
abrupt transition to the relaminarization stage\cite{Avila2023} and
exhibit the exponential decay to laminar flow. The scenario is analogous to the results of DNS of the plane Couette flow\cite{Chantry2014}, in which the streaks' viscous decay is the cause of the exponential decay. In fact, persistent longitudinal large scales are viscously dampened because small eddies decay first and quickly, leaving them without a self-sustaining mechanism when the small eddies transfer energy from the mean shear flow. The distinction is that puffs have a persistent, weakly Reynolds-dependent size that is a characteristic length. Consequently, the presence of a weakly Reynolds-dependent rate of decay is implied\cite{KhanArogetiYakhot2024}. We define the onset of relaminarization based on the phase-space projection of typical chaotic trajectories of the energy of streamwise velocity fluctuations, $e_z$, and the total energy dissipation, $\cal {D}$, shown in figure
\ref{fig: ez-t-ez-vs-D-250last-Re1880-1900-1920} (right). Namely, the point at which a chaotic trajectory abruptly begins to follow a linear trend ($e_z \propto\cal {D}$) is identified as the onset of relaminarization.

\begin{figure}
\centerline{
  \hbox{
    \resizebox{50mm}{!}
    {\includegraphics[width=0.7 \textwidth]{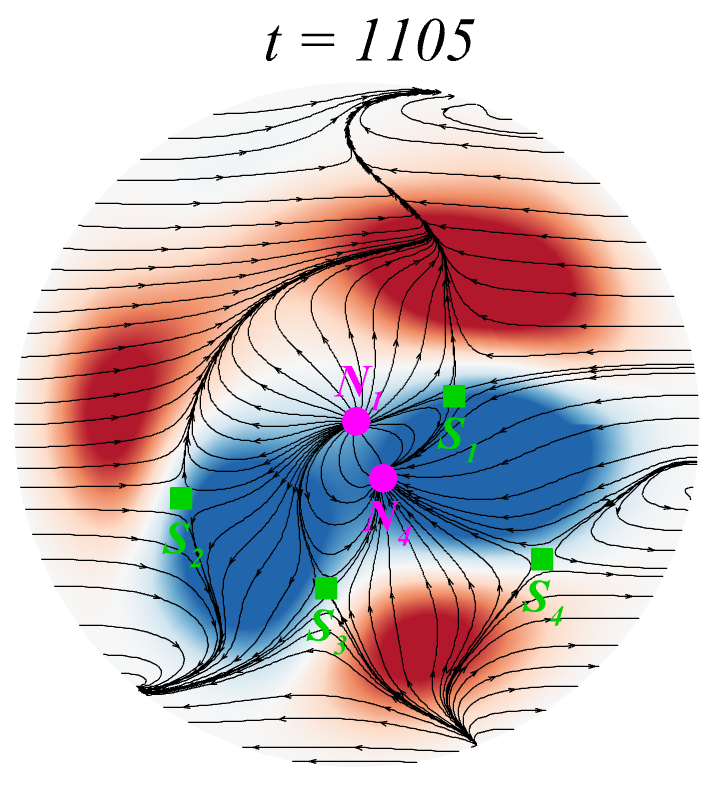}}
    \hspace{1mm}
    \resizebox{50mm}{!}
    {\includegraphics[width=0.7 \textwidth]{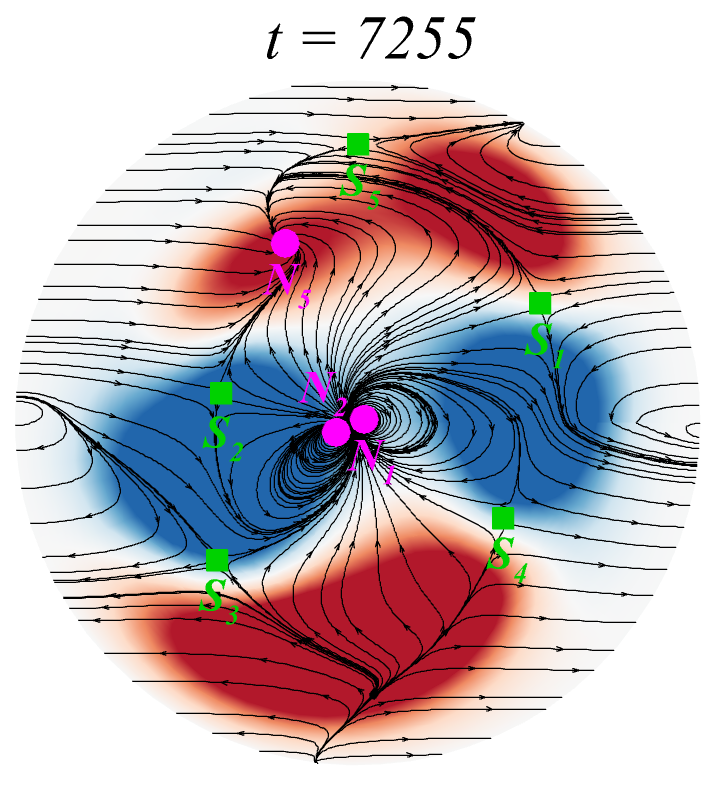}}
    \hspace{1mm}
  }
}
\centerline{
  \hbox{
    \resizebox{50mm}{!}
    {\includegraphics[width=0.7 \textwidth]{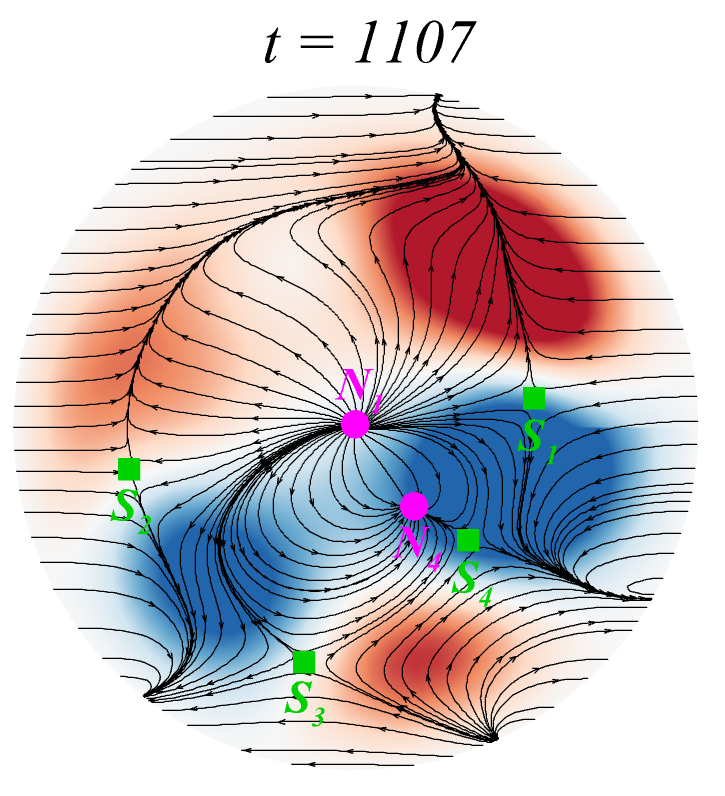}}
    \hspace{1mm}
    \resizebox{50mm}{!}
    {\includegraphics[width=0.7 \textwidth]{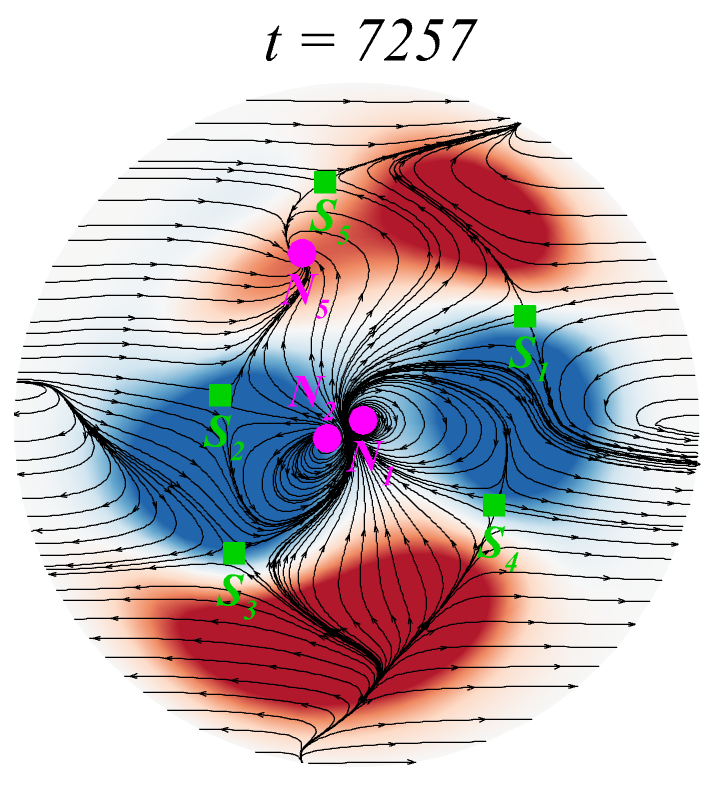}}
    \hspace{1mm}
  }
}
\centerline{
  \hbox{
    \resizebox{50mm}{!}
    {\includegraphics[width=0.7 \textwidth]{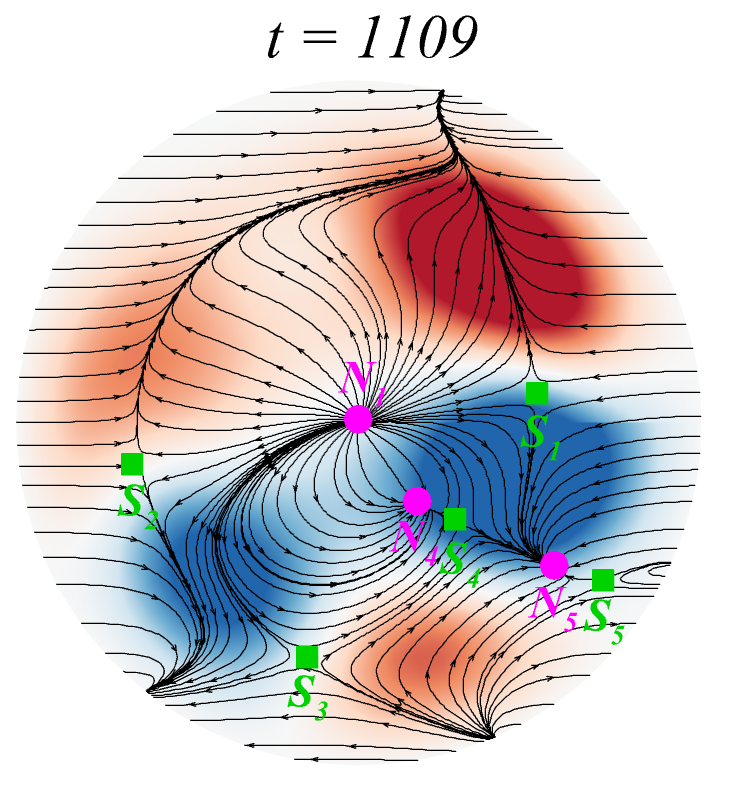}}
    \hspace{1mm}
    \resizebox{50mm}{!}
    {\includegraphics[width=0.7 \textwidth]{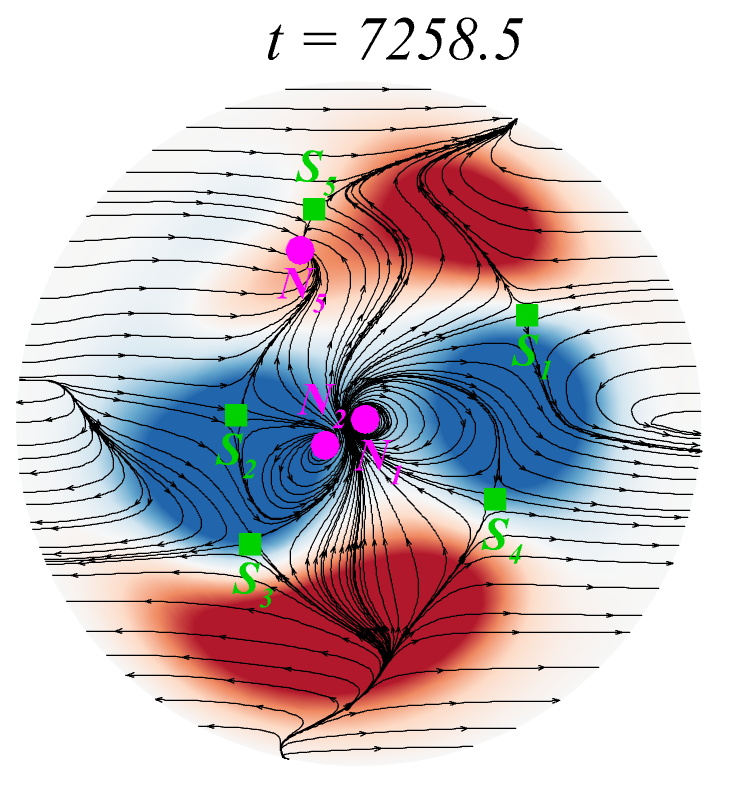}}
  }
}
    \caption{
$Re=1900$, $z=-4D$. Sectional streamlines ($u_r, u_\theta$) for two different initial conditions, IC1 (left) and IC2 (right)
at different
time instants during the last stage of relaminarization; $S_i$ and $N_i$ denote pairs of saddle and nodal points. The onset of relaminarization: $t=1089$ (left) and $t=7240$).
}
        \label{fig: S-N-streamlines-Re1900-IC1-IC2}
\end{figure}

\begin{figure}
\centerline{
  \hbox{
    \resizebox{7.0cm}{!}
    {\includegraphics[width=1.0 \textwidth]{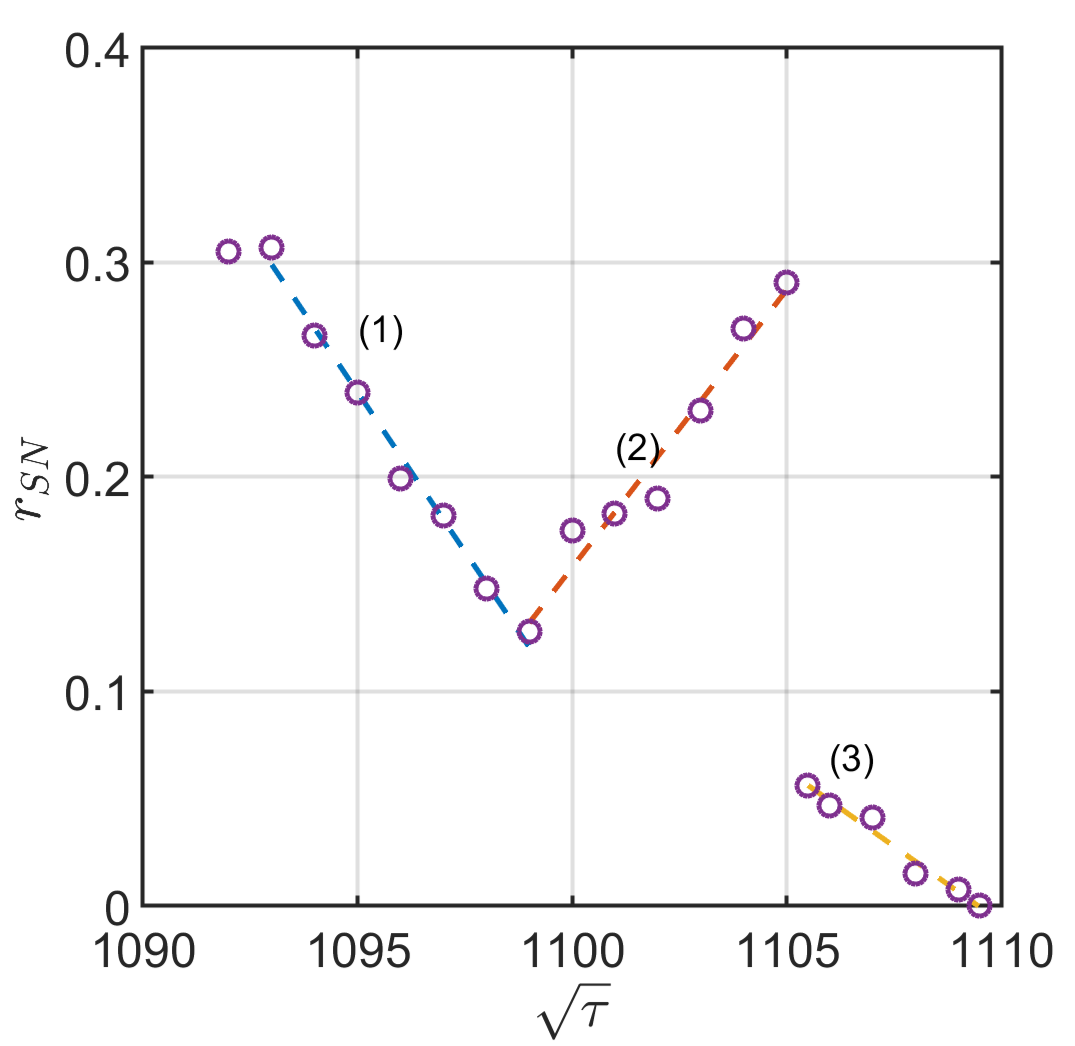}}  
    \hspace{1mm}
    \resizebox{7.0cm}{!}
    {\includegraphics[width=1.0 \textwidth]{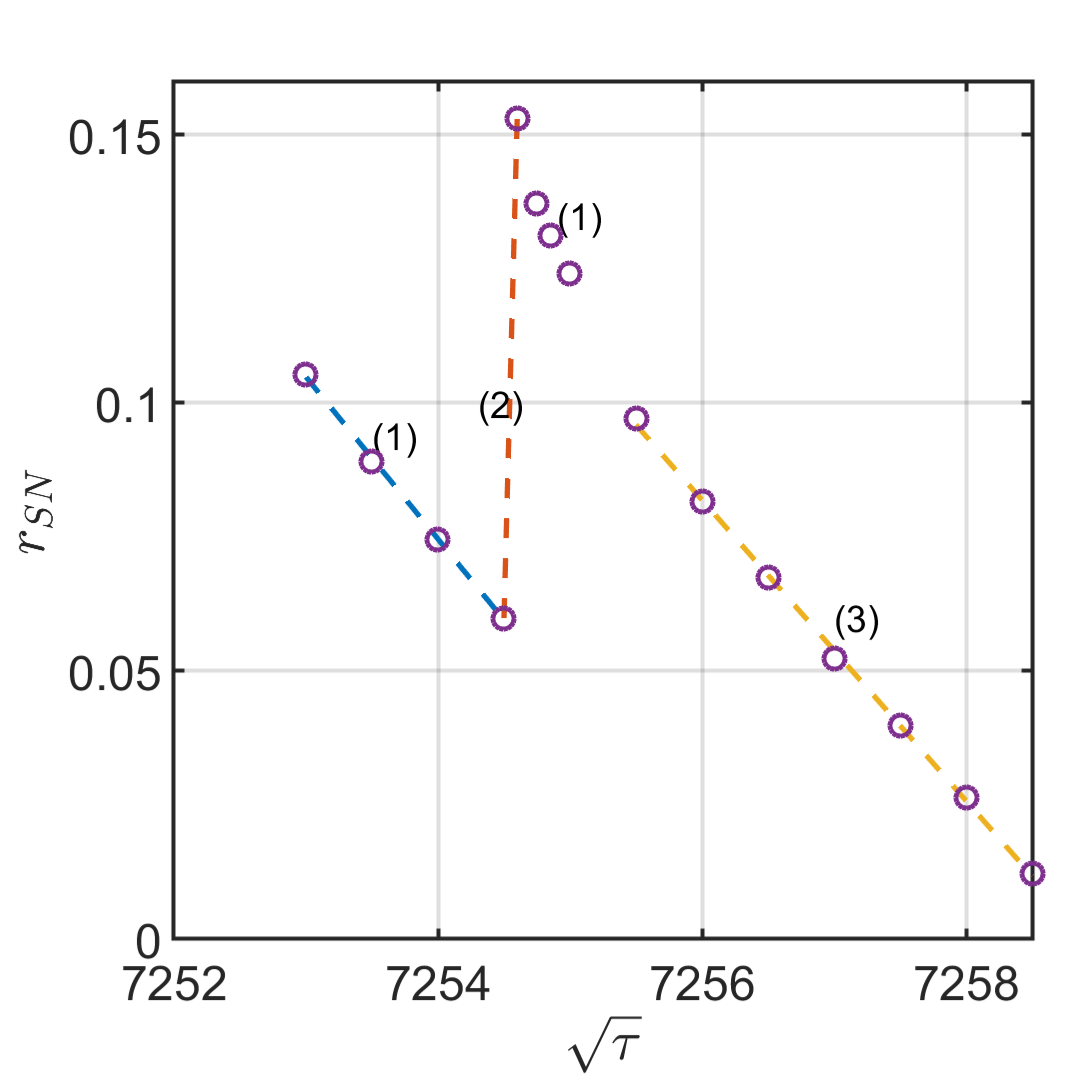}} 
    \hspace{1mm}
  }
}
    \caption{$Re=1900$, $z=-4D$. The saddle-node annihilation during the relaminarization shown in Fig. \ref{fig: S-N-streamlines-Re1900-IC1-IC2}. The square-root scaling, $r_{SN} \propto (\tau_*-\tau)^{1/2}$; $r_{SN}$ is the distance between the saddle and node points. Three distinct stages for the collision of saddles and nodes: approach (1), repulsion (2) and annihilation at the final stage (3).
    }
        \label{fig: sqrt-rSN-vs-t-Re1900-IC1-IC2}
\end{figure}

\begin{figure}
\centerline{
  \hbox{
    \resizebox{70mm}{!}
    {\includegraphics[width=1.0 \textwidth]{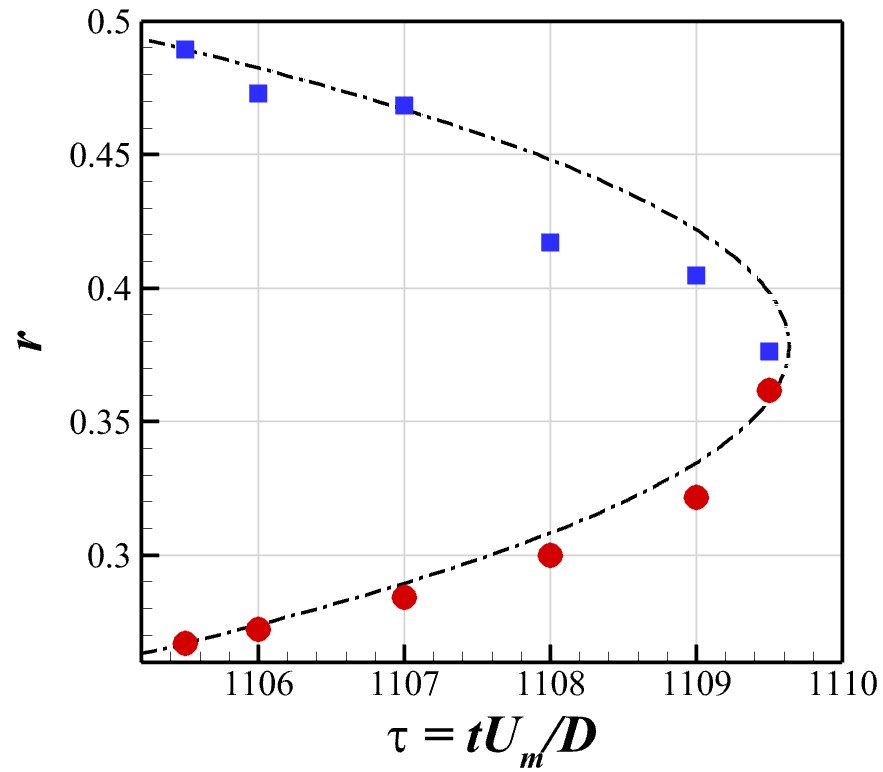}}
    \hspace{1mm}
    \resizebox{70mm}{!}
    {\includegraphics[width=1.0 \textwidth]{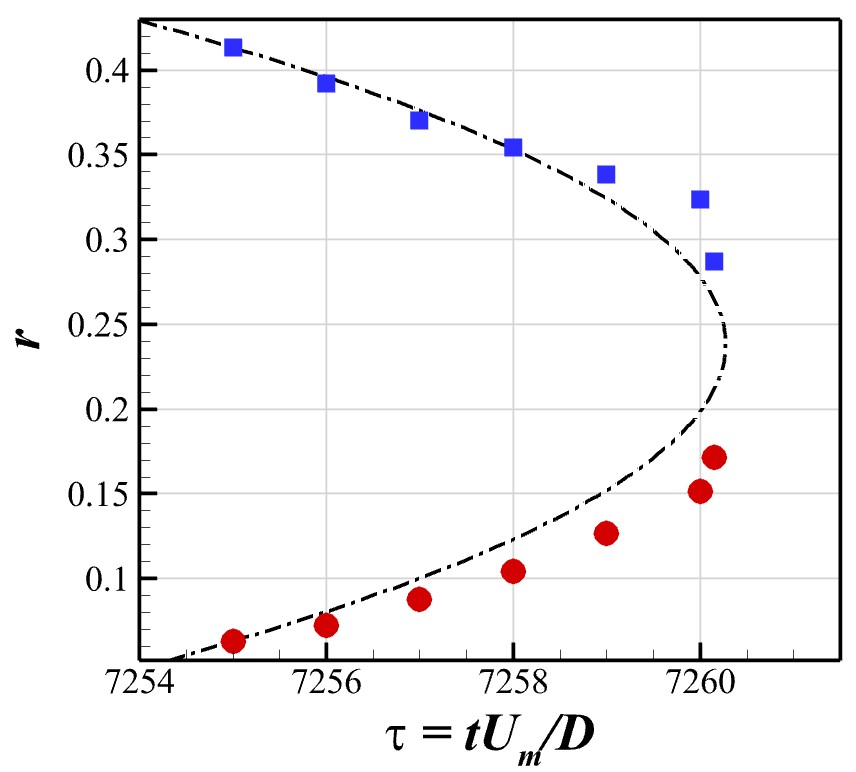}}
    \hspace{1mm}
  }
}
    \caption{$Re=1900$, $z=-4D$. The saddle-node annihilation during the relaminarization shown in Fig.
    \ref{fig: S-N-streamlines-Re1900-IC1-IC2}. The $S/N$ pairs move along the parabola $t_a-t \propto (r-r_a)^2$, $r$ is
    the $r$-coordinate of the saddle and node points, and $t_a$ and $r_a$ are the time and coordinate, respectively, of the annihilation;
    the left and right parabolas correspond to Fig. \ref{fig: S-N-streamlines-Re1900-IC1-IC2}, the $S_4/N_4$ pair (left)
    and  the $S_5/N_5$ pair (right), respectively.
}
        \label{fig: r-vs-t-DNS-Re1900-IC1-IC2}
\end{figure}

\begin{figure}
\centerline{
  \hbox{
    \resizebox{70mm}{!}
    {\includegraphics[width=0.9 \textwidth]{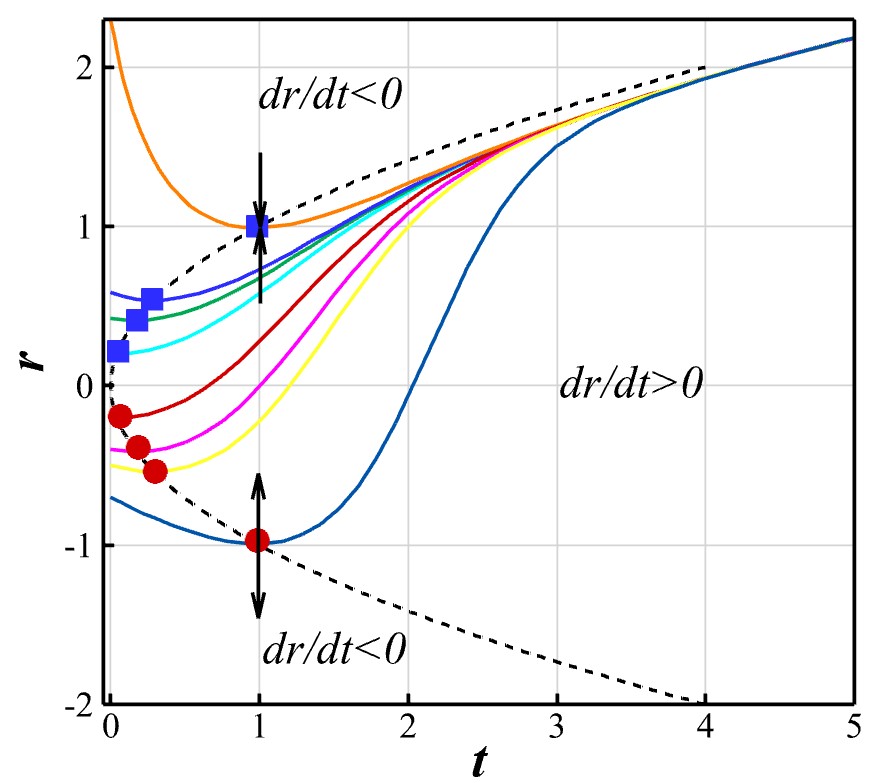}}
    \hspace{1mm}
    \resizebox{70mm}{!}
    {\includegraphics[width=0.9 \textwidth]{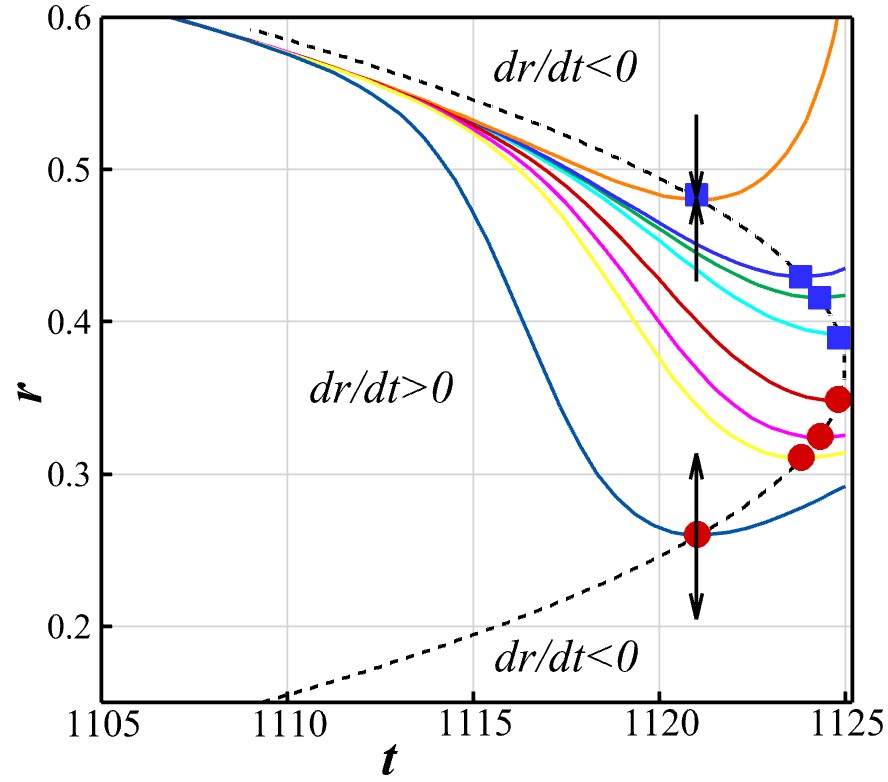}}
    \hspace{1mm}
  }
}
    \caption{(left) The Riccati-type saddle-node bifurcation model, $dr/dt=t-r^2$. The trajectories $r(t)$ are solutions for different initial conditions $r(0)$. (right) The parabola $t_a-t \propto (r-r_a)^2$ is the mirror transformation of that on the left; $r_a$ and $t_a$ are the coordinate and time, respectively, of annihilation; the time-axis ticks are symbolic.
}
        \label{fig: t=r2-phase-curve}
\end{figure}

We study the organized flow structures, focusing on the near-wall streaks and the ($u_r,~u_\theta$) sectional streamlines at the transition region, $z = -4D$.
Figure \ref{fig: S-N-streamlines-Re1900-IC1-IC2} offers an additional perspective on the flow structure following the abrupt exit of a chaotic attractor (turbulence) and the subsequent transition to laminar flow. Analysis of the in-plane sectional
streamline patterns revealed the presence of saddle and nodal points in the flow field. Similar findings have been reported near the turbulent/non-turbulent interface when analyzing the sectional streamlines topology of the velocity fields of a turbulent wake behind a flat plate in the intermittent zone\cite{BissetHuntRogers2002}. In the in-plane velocity field, a node point is a stagnation point where streamlines converge toward (or diverge away from) the point more uniformly, whereas a saddle point is a stagnation point where streamlines converge along one direction and diverge along another\cite{perry1987description}. From figure \ref{fig: S-N-streamlines-Re1900-IC1-IC2}, we  identify two saddle-node pairs that become closer over time,
$S_4/N_4$ (left) and $S_5/N_5$ (right), until they collide and disappear. We note that saddle-node pairs were identified at the interface of the laminar-turbulent transition, $z=-4D$.
The described behavior resembles a saddle-node bifurcation developing over time, ending in annihilation. The topological collapse is not sudden; it develops gradually and adheres to a consistent trajectory in phase-space and physical-space, suggesting a saddle-node bifurcation in the dynamical systems context. In figure
\ref{fig: sqrt-rSN-vs-t-Re1900-IC1-IC2}, we show the distance between a saddle point and a nodal point, identified as a pair, as a function of the square root of the DNS-time of the corresponding snapshot. Straight lines indicate the {\it square-root scaling law}, a generic characteristic of the saddle-node bifurcation\cite{Strogatz2018}. This provides evidence that the disappearance of turbulence-supporting structures occurs via a bifurcation mechanism, rather than being purely stochastic.
An ordinary differential equation $dr/dt=\mu - r^2$ ($\mu>0$ is a constant) is a classic example of such a system with the saddle-node bifurcation. It has two fixed points: $r_1=\sqrt{\mu}$ (stable) and $r_2=-\sqrt{\mu}$ (unstable). As $\mu$ goes down, the two fixed points get closer until they hit each other at $\mu=0$ and disappear, indicating annihilation.
In the context of puff relaminarization in this study, the square-root scaling of the distance between the saddle and node points, $r_{SN} \propto \sqrt{\tau}$, means that a $S/N$ pair moves along a parabola, as illustrated in figure \ref{fig: r-vs-t-DNS-Re1900-IC1-IC2}, where the $r$-location of the $S/N$ pairs of snapshots of different times are projected on
a ($r, \tau$) parabolic curve.\\
\noindent Based on the parabola in figure \ref{fig: r-vs-t-DNS-Re1900-IC1-IC2}, we propose a straightforward yet illustrative model of the saddle-node bifurcation:
\begin{equation}\label{eq: drdt=t-r2}
  \frac{dr}{dt}=t-r^2,
\end{equation}
where $r$ is the $r$-coordinate of the saddle and node points. Equation (\ref{eq: drdt=t-r2}) is the Riccati equation. The Riccati equation can be transformed into a second-order linear differential equation, known as the Airy equation
\footnote{Substituting $r=\dot u/u$ into Eq. \ref{eq: drdt=t-r2} yields the second order Airy equation\cite{AiryBook2010} for $u(t):~\ddot{u}-tu=0$. Its general solution reads: $u(t)=C_1Ai(t)+C_2Bi(t)$, where $Ai(t)$ and $Bi(t)$ are the Airy functions of the first and second kind, respectively.}.
While we cannot draw a standard phase line because Eq. (\ref{eq: drdt=t-r2}) is not autonomous, analyzing the dynamics in the $(r,t)$-plane is very beneficial. In Figure \ref{fig: t=r2-phase-curve} (left), the plane is divided into two regions by the parabola $t=r^2$, where the lower branch is unstable and the upper branch is stable.
\noindent The trajectories $r(t)$ are solutions to Eq. (\ref{eq: drdt=t-r2}) for different initial conditions $r(0)$. These trajectories represent the projection of $S/N$ pairs of snapshots from distinct times onto a single $(r, t)$ plane. They cross the parabola, indicating the nodal (upper branch) and saddle (lower branch) points. Lastly, using a mirror transformation, we bring the parabola to the shape obtained by the DNS, as shown in figure \ref{fig: t=r2-phase-curve} (right). Therefore, the Airy equation, which arises in the analysis of dynamics near critical points in physics phenomena\cite{AiryBook2010}, leaves its signature by describing the saddle-node bifurcation that is observed during the relaminarization of a turbulent puff. Specifically,
the number of saddle and nodal points decreases gradually during the relaminarization in the near-wall region. The DNS results in certain instances clearly demonstrate the convergence of saddle and node points, which ultimately leads to their annihilation.\\
{\bf Acknowledgement.} This study was supported by the Israel Science Foundation Grant 2228/22.
\bibliography{refspuff2.bbl}

\end{document}